\documentclass[pdftex,12pt,a4paper]{article}
\usepackage[squaren]{SIunits}
\usepackage{amssymb}
\usepackage[pdftex]{graphicx}
\usepackage{fancyhdr}
\usepackage{enumerate}
\usepackage{amsmath}
\usepackage{color}
\usepackage{setspace}
\usepackage[latin1]{inputenc}
\usepackage[pdftex,bookmarks=true]{hyperref}
\usepackage{cite}
\usepackage{authblk}

\begin{document}


\title{
Sealed operation, and circulation and purification of gas in the HARPO TPC
}

\author[1]{M.~Frotin\thanks{\noindent corresp. authors: frotin@llr.in2p3.fr, philippe.gros@llr.in2p3.fr}}
\author[1]{P.~Gros}
\author[2]{D.~Atti\'e} 
\author[1]{D.~Bernard} 
\author[3]{V.~Dauvois} 
\author[2]{A.~Delbart} 
\author[3]{D.~Durand} 
\author[1]{Y.~Geerebaert} 
\author[3]{S.~Legand} 
\author[2]{P.~Magnier} 
\author[1]{P.~Poilleux} 
\author[1]{I.~Semeniouk}

\affil[1]{LLR, Ecole Polytechnique, CNRS/IN2P3, 91128 Palaiseau, France}
\affil[2]{IRFU, CEA Saclay, 91191 Gif-sur-Yvette, France}
\affil[3]{DEN/DANS/DPC/SECR/LRMO, CEA Saclay, 91191 Gif-sur-Yvette, France}

\maketitle

\abstract{
  HARPO is a time projection chamber (TPC)
  demonstrator of a gamma-ray telescope and polarimeter in the MeV-GeV
  range, for a future space mission. 
  We present the evolution of the TPC performance over a five month
  sealed-mode operation, by the analysis of
  cosmic-ray data, followed by the fast and complete recovery of the
  initial gas properties using a lightweight gas circulation
  and purification system. 
}


\section{Introduction}
\label{sec:introduction}

High-energy astronomy is hindered by the sensitivity gap between the
energy ranges over which the Compton and the pair telescopes are
efficient, making the $\gamma$-ray sky in the range 0.1 -- 100\,MeV
almost unexplored.
Furthermore no polarimeter with a decent sensitivity above 1\,MeV has
ever been flown in space.

On the low-energy side of the gap, a number of projects are aiming at
 an improved sensitivity with respect to COMPTEL.
Of these, those using stacks of silicon wafers without any tungsten
converter can also detect pair conversions 
\cite{ASTROGAM:2015,Wu:2014tya,Moiseev:2015lva}
with an angular resolution of $\approx 1.5^\circ$ @ 100\,MeV,
a significant improvement over that of the Fermi/LAT 
($\approx 6^\circ$ @ 100\,MeV \cite{Ackermann:2012kna}).
These stacks can have some sensitivity to polarisation if
the wafers are thin enough and provided that matching can be performed
in the first wafer after the conversion wafer, i.e. that each of the
hits in the $x$ direction can be assigned to one of the two hits in the
orthogonal ($y$) direction. In subsequent wafers, the azimuthal information
carried by the pair is washed out by multiple scattering
\cite{ASTROGAM:pol:2015}.

In the quest for even thinner wafers, it is tempting to use an even less
 dense detector such a gas detector. 
In the case of an optimal tracking in the presence of multiple
 scattering such as with a Kalman filter, an even better angular resolution
 of $\approx 0.4^\circ$ @ 100\,MeV can be
 achieved \cite{Bernard:2012uf} providing a point-like source
 sensitivity as low as
$ 10^{-6} \mega\electronvolt/(\centi\meter^2 \second)$
 @ 100\,MeV with a $10\,\kilo\gram$ gas time projection chamber (TPC),
while that of COMPTEL plateaued at 
 $\approx 10^{-4} \mega\electronvolt/(\centi\meter^2 \second)$ 
\cite{Schoenfelder}.
In such gas detectors the dilution of the azimutal information carried by
 the pair is close to unity, 
so that the
sensitivity to polarisation is excellent \cite{Bernard:2013jea}.
Finally and most importantly 
track matching can easily be performed by comparing the distribution
of the energy deposited along the tracks 
in the $x$ and $y$ directions thanks to the large fluctuation in the
local deposition of energy along the track \cite{Pisa2012}.
These unique features come at the cost of the need to maintain a 
good-quality gas in which the collection of the ionizing electrons
is not affected by impurities with a large electron attachment.
For example the EGRET gas
(99.5\% neon, 0.25\% argon, and 0.25\% ethane) became contaminated by
the breakdown of ethane by spark chamber firings and to a small degree by
residual outgassing.
The five planned gas exchanges were carried out once per year, after
which the lack of fresh gas became a serious issue \cite{EGRET:1999}.
These concerns have triggered the development of detectors built out of
 low outgassing rate materials.
For example, for the GEMS X-ray polarimeter TPC, the design of which
involves copper clad liquid crystal polymer (LCP) gas electron
multipliers (GEM), polyetheretherkeytone (PEEK) dielectric structure
on a gold-plated titanium frame, a 23 year lifetime is expected
\cite{GEMS:JHill:SPIE:8859}.

We report here on a gas purity study performed with the HARPO detector, a TPC
prototype of a $\gamma\rightarrow e^+e^-$ telescope built with
techniques much better suited to our actual budget and that we have
 studied in the laboratory with radioactive sources
\cite{Gros:TIPP:2014} and with cosmic rays \cite{Bernard:2014kwa}, and
that we have tested recently \cite{Shaobo} in a $\gamma$-ray beam.

\section{Description of the HARPO TPC}
\label{sec:description:TPC}

\begin{figure} [th]
 \begin{center} 
 \includegraphics[width=1\linewidth]{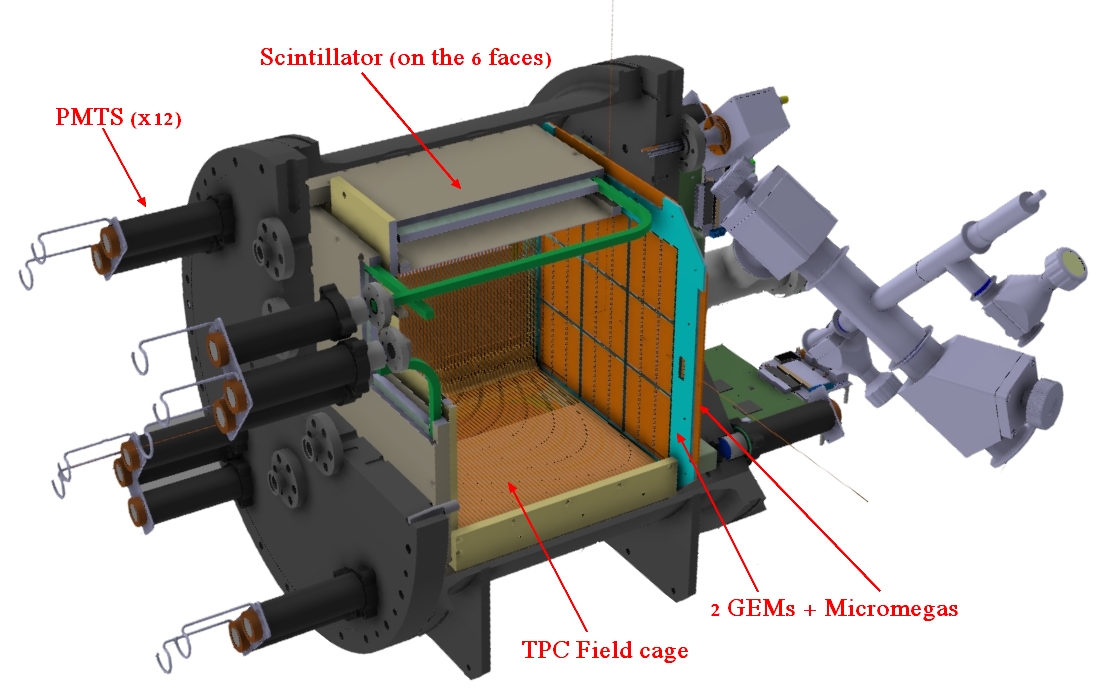}
\caption{\label{fig:demonstrator}
A 3D CAD drawing of the HARPO demonstrator.}
\end{center}
\end{figure}

HARPO is a 30\,cm cubic TPC surrounded by 6 scintillator plates, each equipped with two wavelength shifter bars and two photomultipliers (PMTs) for trigger and background rejection.
The gas container is an aluminum cylinder designed to be operated from few mbar up to 5\,bar in sealed mode (Fig.~\ref{fig:demonstrator}). 
With a mixture of 95\% of argon 5.0 and 5\% of iC$_4$H$_{10}$ 3.5 and at a 220\,V/cm drift electric field, the electron drift velocity is about $3\,\centi\metre/\micro\second$.
The TPC end-plate is a "hybrid" MicroPattern Gas Detector (MPGD)
 composed of two GEM foils\cite{SAULI:1997} with 2\,mm
spacing above a $128\,\mu$m amplification gap
bulk-micromegas\cite{bulk}.
The anode collection plane is segmented in $2 \times 288$ strips in the
two orthogonal $x, y$ directions and the signal is read by an electronic chain composed of
two T2K/TPC Front-End cards\cite{AFTER}, two FEMINOS back-end cards and a Trigger Clock Module~\cite{Feminos}. 
A simple coincidence between scintillators is used here for triggering on cosmic rays.

\section{Description of the circulation and purification system (CAPS)}
\label{sec:description:CAPS}

\begin{figure} [th]
 \begin{center} 
 \includegraphics[width=0.29\linewidth]{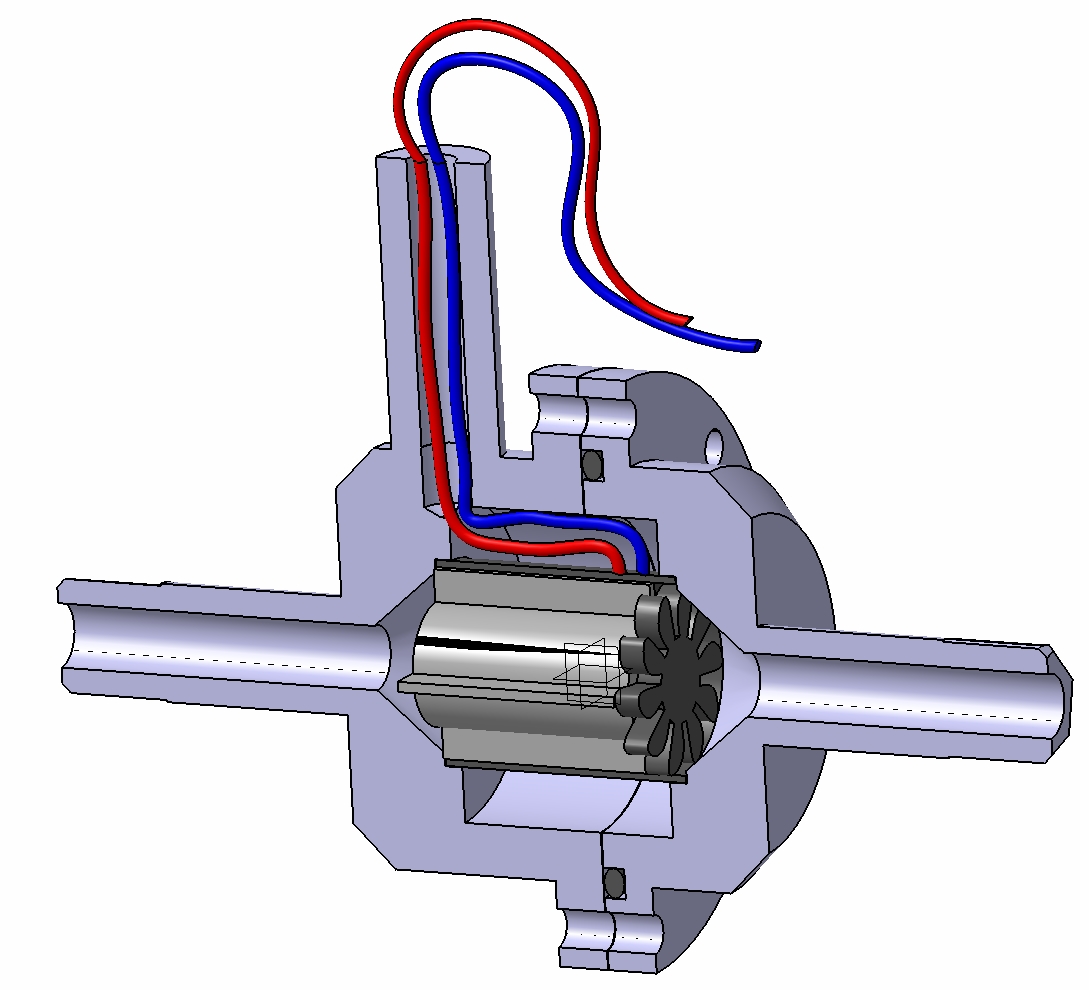}
\hfill
 \includegraphics[width=0.7\linewidth]{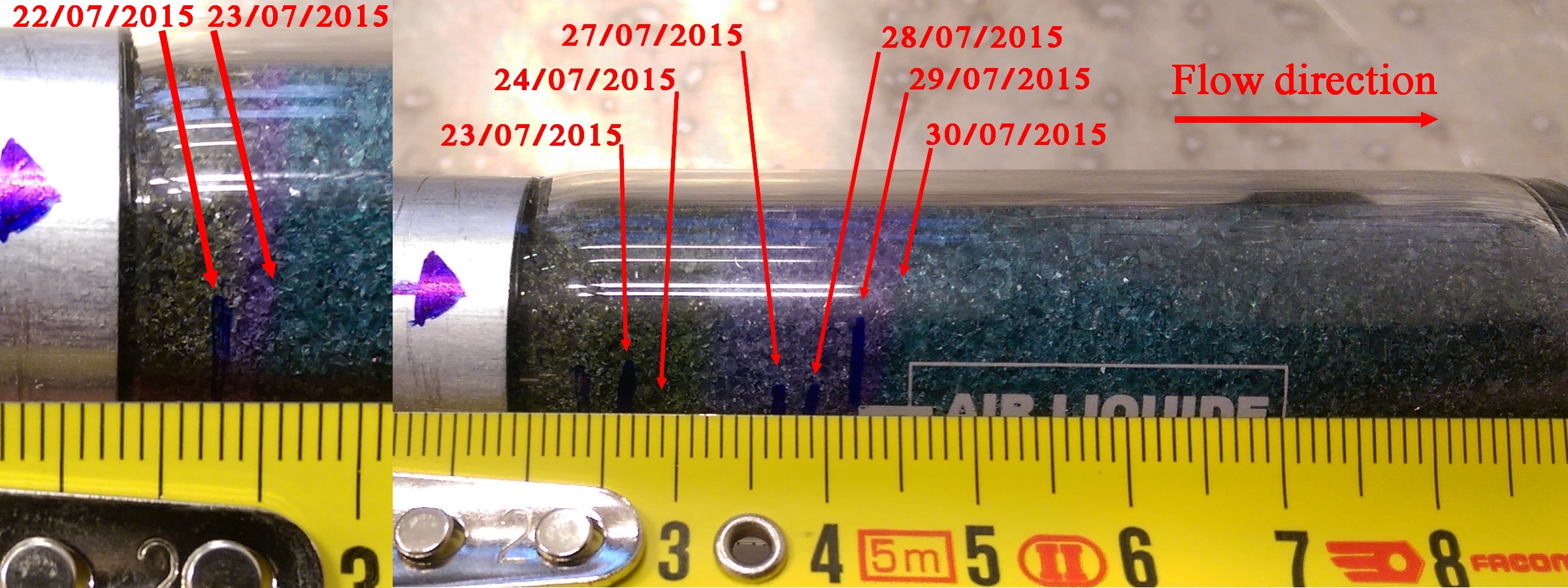}
 \caption{\label{fig:Circulator}
 (Left) Sectional view of the circulator (CEA patent BD15730SG).
 (Right) Picture of the Oxisorb cartridge after 10 days of continuous use.
}
 \end{center}
\end{figure}

\begin{figure} [th]
 \begin{center} 
 \includegraphics[trim=0 10 0 10,clip,width=0.9\linewidth]{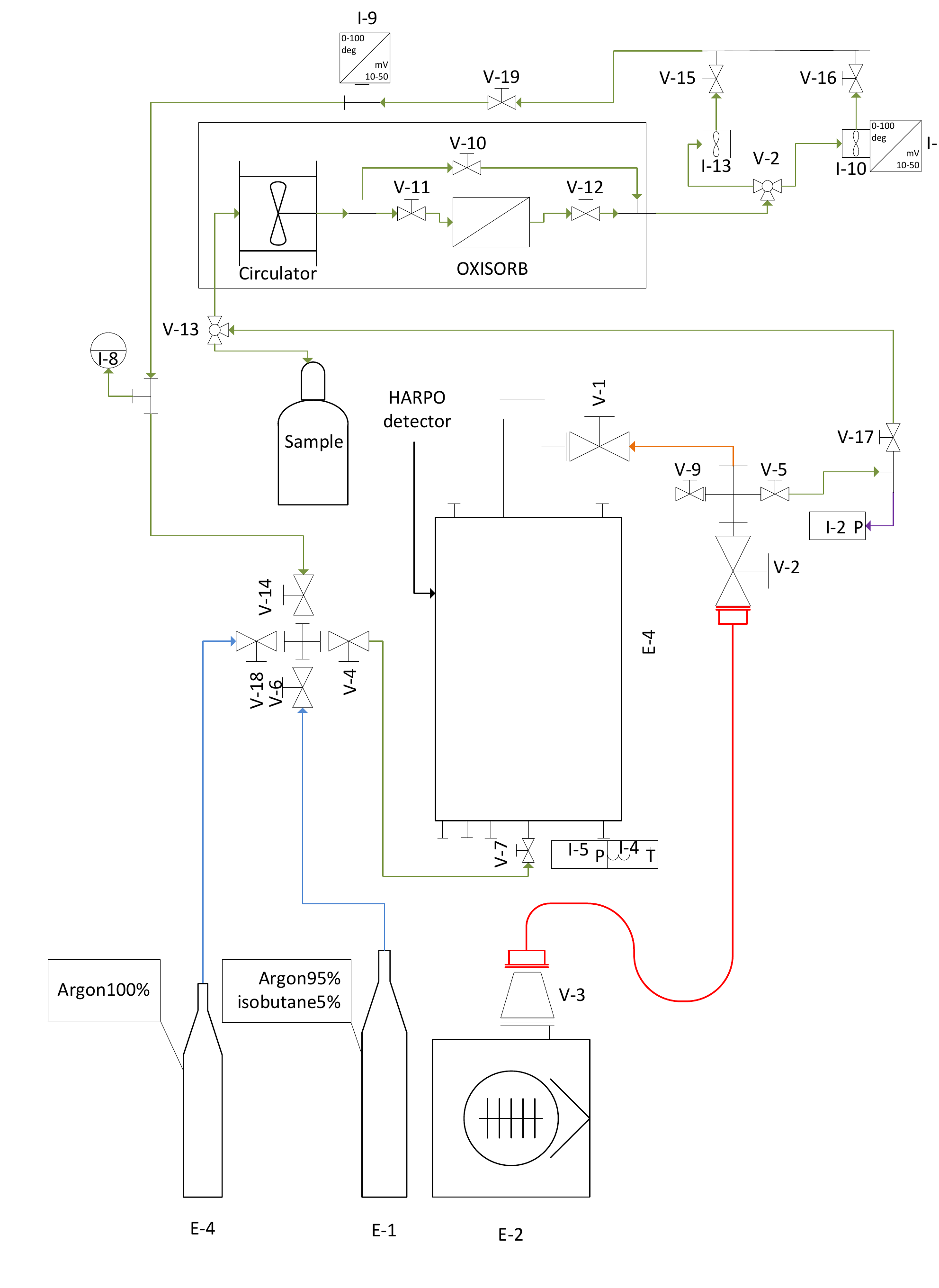}
 \caption{\label{fig:layout:of:the:gas:system}
 Layout of the gas system.}
 \end{center}
\end{figure}

We used a circulator composed of a simple turbine 
(Fig.~\ref{fig:Circulator}, left).
The gas was lead through an Oxisorb purification cartridge that can absorb up to $0.5~\liter$ H$_2$O and $0.1~\liter$ O$_2$, with output purity as low as $<$~5~ppb for O$_2$ and 30~ppb for H$_2$O~\cite{oxisorb}.
We measured the gas flow with a F-111B Bronkhorst mass flowmeter (Fig. \ref{fig:layout:of:the:gas:system}, I10) that was calibrated for our gas mix at $2~\bbar$ and $20~\celsius$.
We monitored the gas pressure and temperature with a Keller EV-120/PA(A)-33X sensor (Fig. \ref{fig:layout:of:the:gas:system}, I4 and I5) with ranges of $0 - 50~\celsius$ and $0 - 10\,\bbar$ and resolutions of $0.01\,\celsius$ and $1\,\milli\bbar$.
The outgassing areas inside the vessel amount to $0.82\,\meter^2$~Aluminum, $3.50\,\meter^2$~PVC, $0.92\,\meter^2$~PCB, $1.63\,\meter^2$~PVT, for a gas volume $V_{\textrm{det}}=84~\liter$.

We tested the gas circulation system with a helium leak detector (ASM 310) down to $3.3\cdot10^{-8}~\milli\bbar\cdot\liter/\second$, after which
we opened the connection between the CAPS and the detector. 
This caused a pressure drop from $2015~\milli\bbar$ to $1960~\milli\bbar$ at $27\,\celsius$,
with a flow of $1.5~\liter/\hour$.
%
On July 20 2015, we opened the Oxisorb purificator in vacuum which lead to a new pressure drop to $1932~\milli\bbar$ when the connection to the detector was opened.
The CAPS then ran continuously, with a flow through the filter of $f_{\textrm{circ}} = 1.4~\liter/\hour$.
In these conditions
we expect the water and oxygen content in the gas, $C$, to evolve
following a law of the form:
\begin{equation}
C = C_{0}e^{-t/\tau_{\textrm{purif}}} + K_{\textrm{cont}}\tau_{\textrm{purif}},
\end{equation}
with 
$\tau_{\textrm{purif}} = {V_{\textrm{det}}} / {\epsilon_{\textrm{purif}} f_{\textrm{circ}}}$, 
where $C_{0}$ is the original content and
$K_{\textrm{cont}}$ is the contamination rate from leaks and outgassing. 
In case the purification efficiency of the filter
is
$\epsilon_{\textrm{purif}} = 1$, 
the purification rate is equal to $\tau_{\textrm{purif}} = 60~\hour$.

The Oxisorb purificator color ranges from blue to dark brown with the quantity of H$_2$O and O$_2$ absorbed. 
At the beginning of the purification, we observed a purple ring of 3 mm thickness growing $3\,\milli\meter$ per day during the first 4 days.
It stabilised at $12\,\milli\meter$ after a few days (Fig.~\ref{fig:Circulator}).
This indicates that the cartridge was never saturated.

Several gas samples were taken for analysis, which lead to density drops (lines 1, 2, and 3 on Fig. \ref{fig:fullTime}).
Before and after every operation that affected the gas in the vessel, data was taken with the TPC for consistency checks.

\section{Data taking and analysis}
\label{sec:analysis:method}

\begin{figure}[ht]
\begin{center}
 \includegraphics[width=0.7\linewidth]{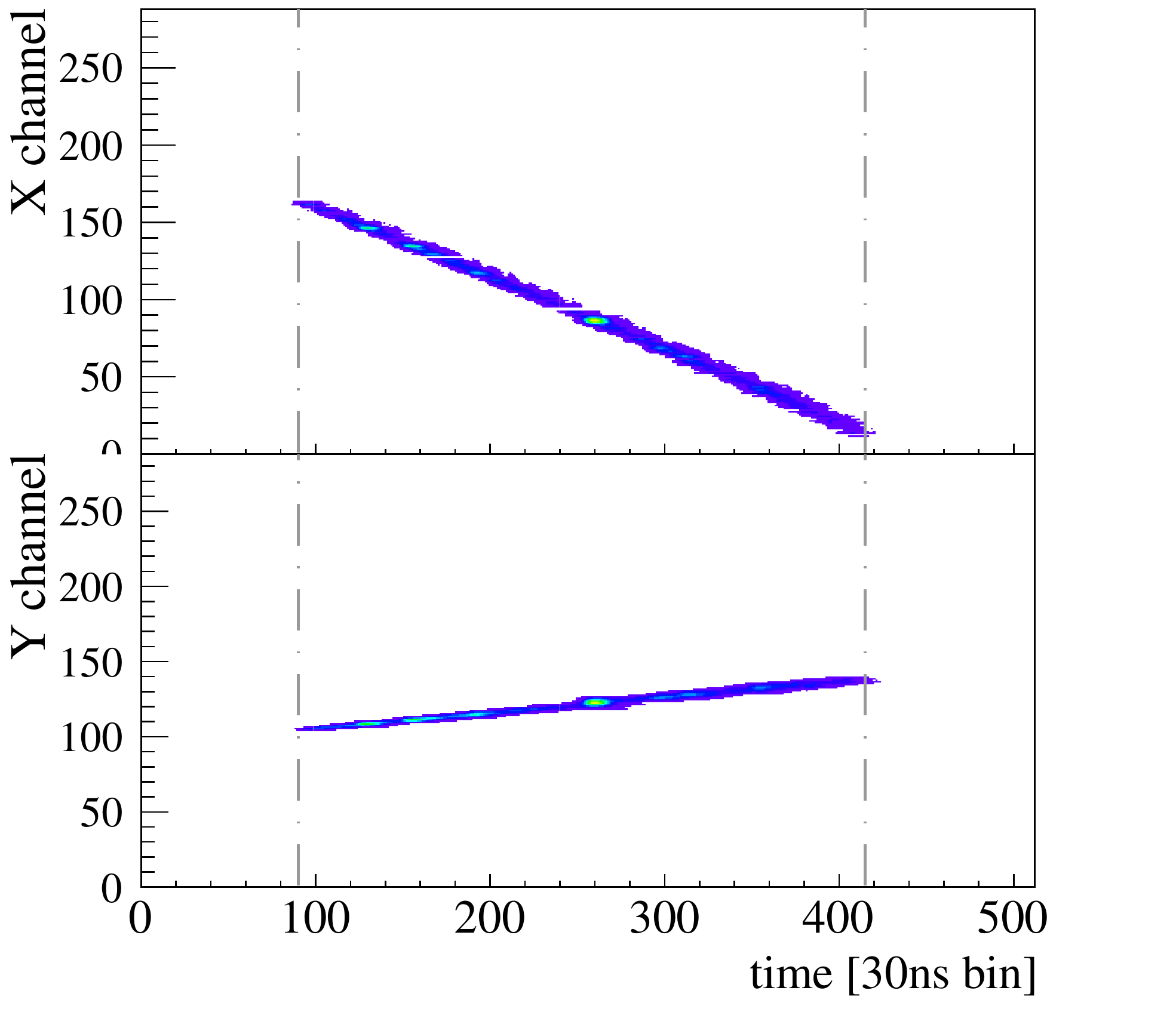}
 \caption{
 A cosmic-ray event in HARPO.
 The ``maps'' show the charge distributions in the projection X-time and Y-time.
 \label{fig:event}
 }
\end{center}
\end{figure}

\begin{figure}[ht]
\begin{center}
 \includegraphics[trim=0 10 0 0,clip,width=0.82\linewidth]{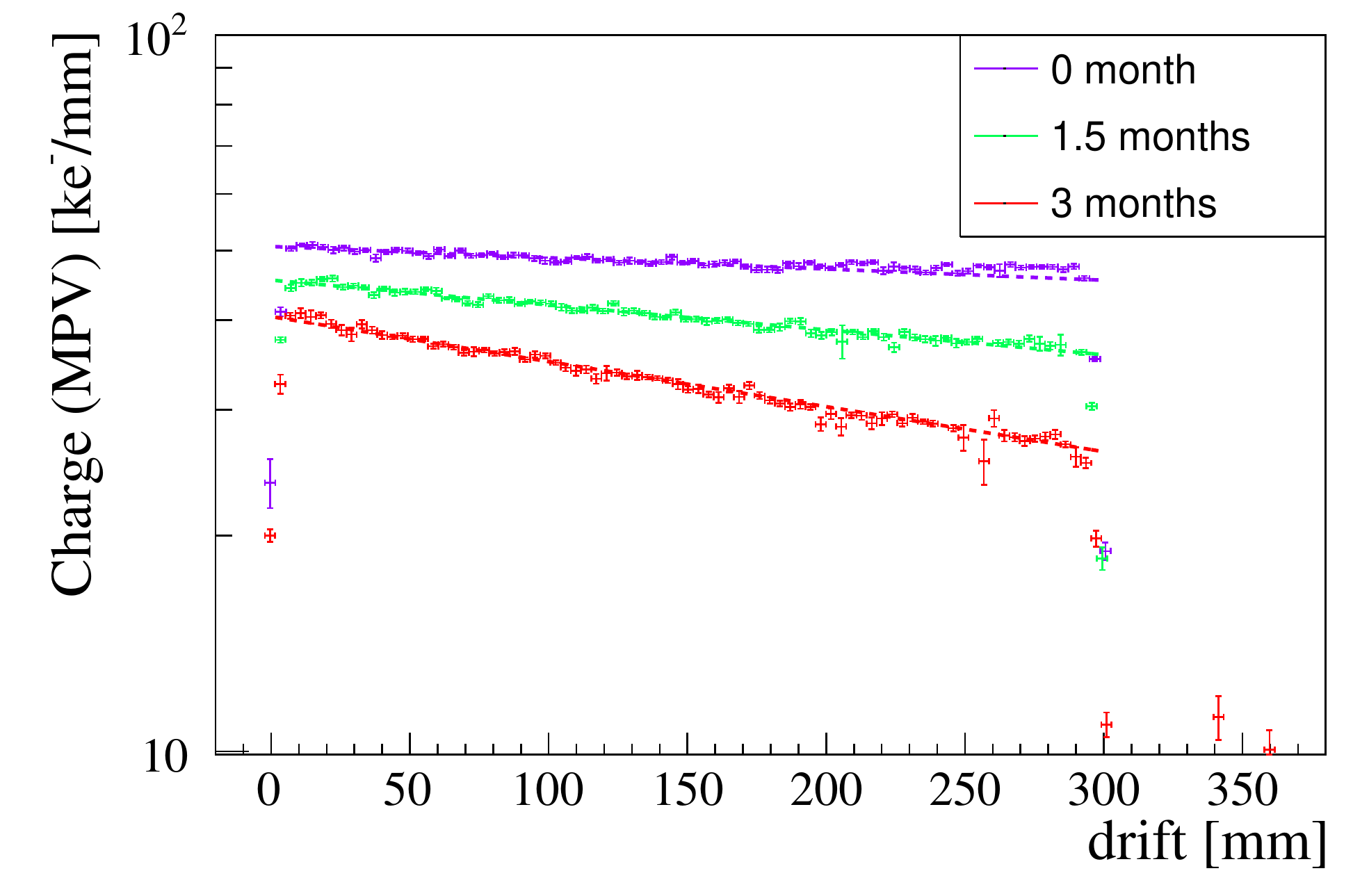}
 \caption{
 Charge as a function of the drift distance, for three runs, 6 weeks apart, normalised to the reference run 2008.
 \label{fig:QrelVsZ}
 }
\end{center}
\end{figure}

We took cosmic-ray data once per week from February to July,
with the detector set so that the drift field was vertical.
Most runs were one hour long, recording a few 10,000~events.
After starting the purification on July 20 we took data more frequently.
The high voltage was only turned on when taking data.
Due to the vertical detector orientation, most of the recorded tracks are approximately aligned with the drift direction.
Figure~\ref{fig:event} shows the $x,t$ and $y,t$ projections for a typical event.
We select events with a single track in each projection, with a high
momentum muon selection based on the $\chi^2$ cut of a linear fit.
For each time bin $t_i$ ($30\,\nano\second \approx 1\,\milli\metre$),
we
obtain the charge deposited by a track, $Q(t_i)$, summing over all the
channels associated with the track.
After an angle-effect correction using the direction $\vec{u}$ of the
track,
\begin{equation}
 \frac{dE}{dx}(t_i) = Q(t_i) \times u_{z}, 
\end{equation}
we obtain the $dE/dx(t_i)$ in
ADC units, that is up to an overall factor that includes the gain of the amplification of
both the MPGDs and that of the readout electronics.
\begin{figure}[ht]
\begin{center}
 \includegraphics[trim=0 10 0 10,clip,width=0.7\linewidth]{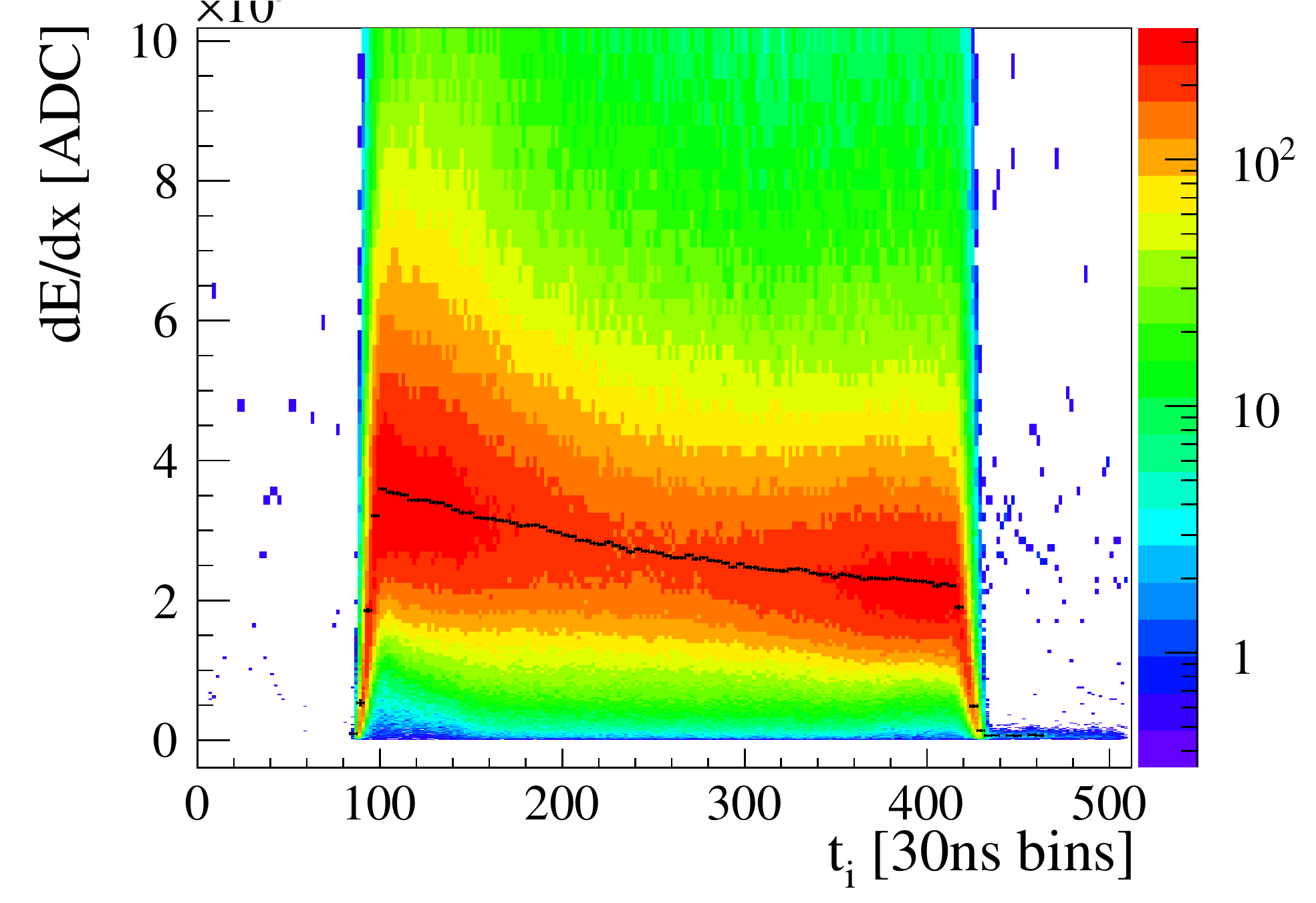}
 \caption{
 Normalised charge as a function of the drift time for a $6000\,\second$ cosmic-ray run.
 \label{fig:QvT}
 }
\end{center}
\end{figure}
The resulting distribution (Fig.~\ref{fig:QvT}) 
 is also affected by threshold and saturation effects.
To mitigate them, we estimate the most probable value (MPV, solid line in Fig.~\ref{fig:QvT}) of the charge for each time bin with a Landau-function fit.
From this distribution we extract : 
\begin{description}
\item[The drift velocity.] 
As the tracks are selected to traverse the full TPC length, 
 the drift-time duration provides a measurement of the drift velocity.

\item[The gain.]
 The height of the distribution at short drift times is proportional to the amplification gain.

\item[The attenuation.]
 The signal loss along the drift from electron capture is visible in Fig.~\ref{fig:QvT}.
 We fit the central part of the distribution, which is less affected by the threshold and saturation effects, with an exponential function.
\end{description}
These measurements are affected by systematic effects.
Since the detector configuration was kept unchanged throughout the
data taking, they cancel in the relative value compared to
a reference run, chosen to be run 2008 (Feb. 18th), i.e. 
right after the fresh gas fill.
Figure~\ref{fig:QrelVsZ} shows the relative charge distribution as a
function of the drift distance for several runs taken at a few weeks
interval.

\section{Results}
\label{sec:results}

\begin{figure}[ht]
\begin{center}
 \includegraphics[trim=0 10 0 0,clip,width=\linewidth]{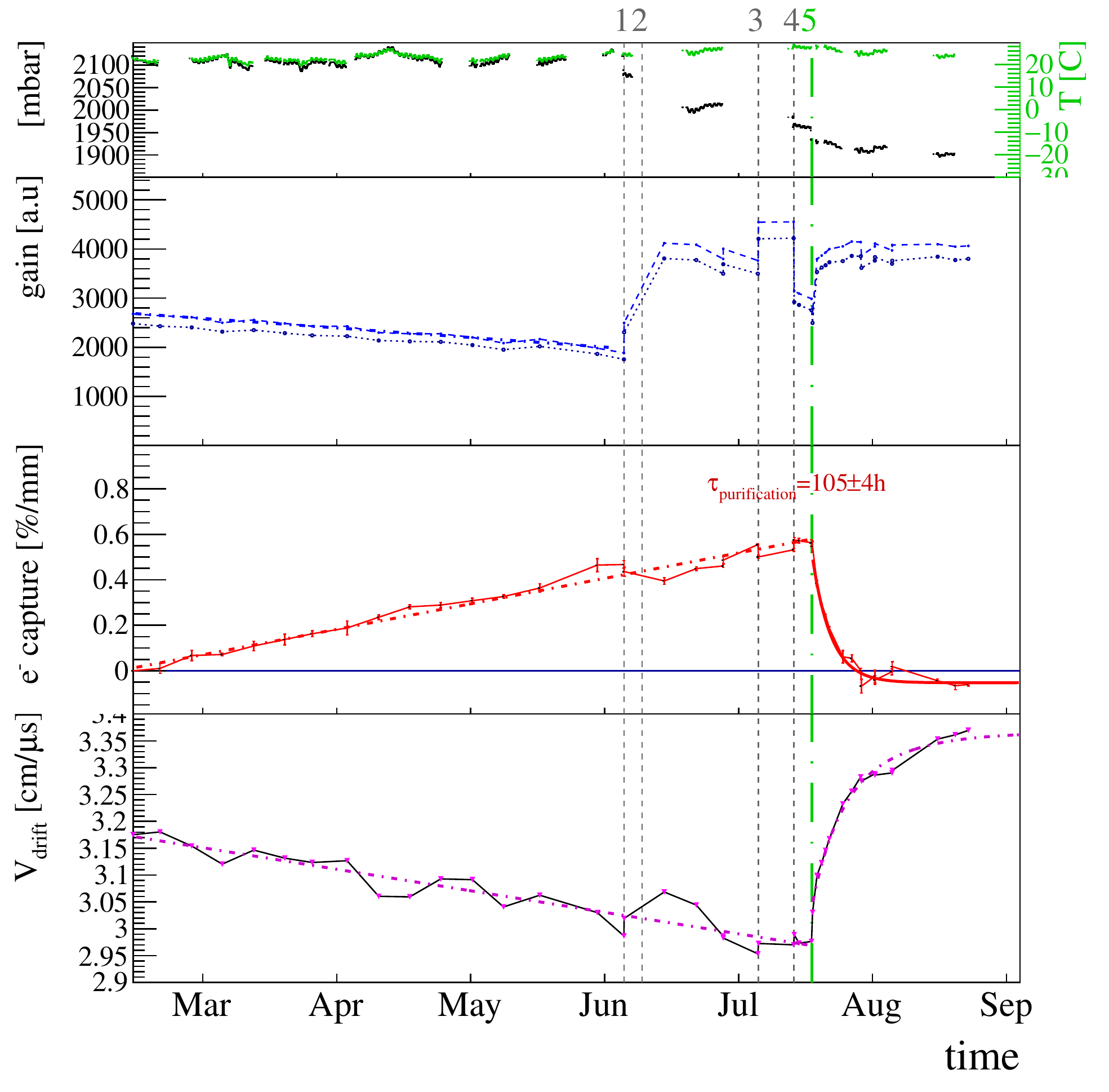}
 \caption{Time evolution 
  of the gain, of the attenuation and of the drift velocity.
 We can see several variations of the gain due to the pressure loss when gas samples were extracted.
 For the first five months during which the TPC was sealed we see a regular degradation of all three parameters.
 On July 20 (line~5), the CAPS was turned on.
 \label{fig:fullTime}
 }
\end{center}
\end{figure}

For the first five months, while the TPC was 
sealed, we observe a steady degradation of the gain, of the drift velocity and of the electron attenuation
(Fig.~\ref{fig:fullTime}).
After
 we turned the CAPS on, 
 the drift velocity and the attenuation got back to their initial values.
We corrected for the gain changes due to the density decrease induced by our extraction of gas
samples and extension of the volume with the CAPS circuit, by
adjusting the MPGD voltages.
The pressure change also affected the drift velocity but to a
negligible extent as shown by Garfield simulations\cite{Garfield},
compared to the observed effect.

\begin{figure}[ht]
\begin{center}
 \includegraphics[width=0.92\linewidth]{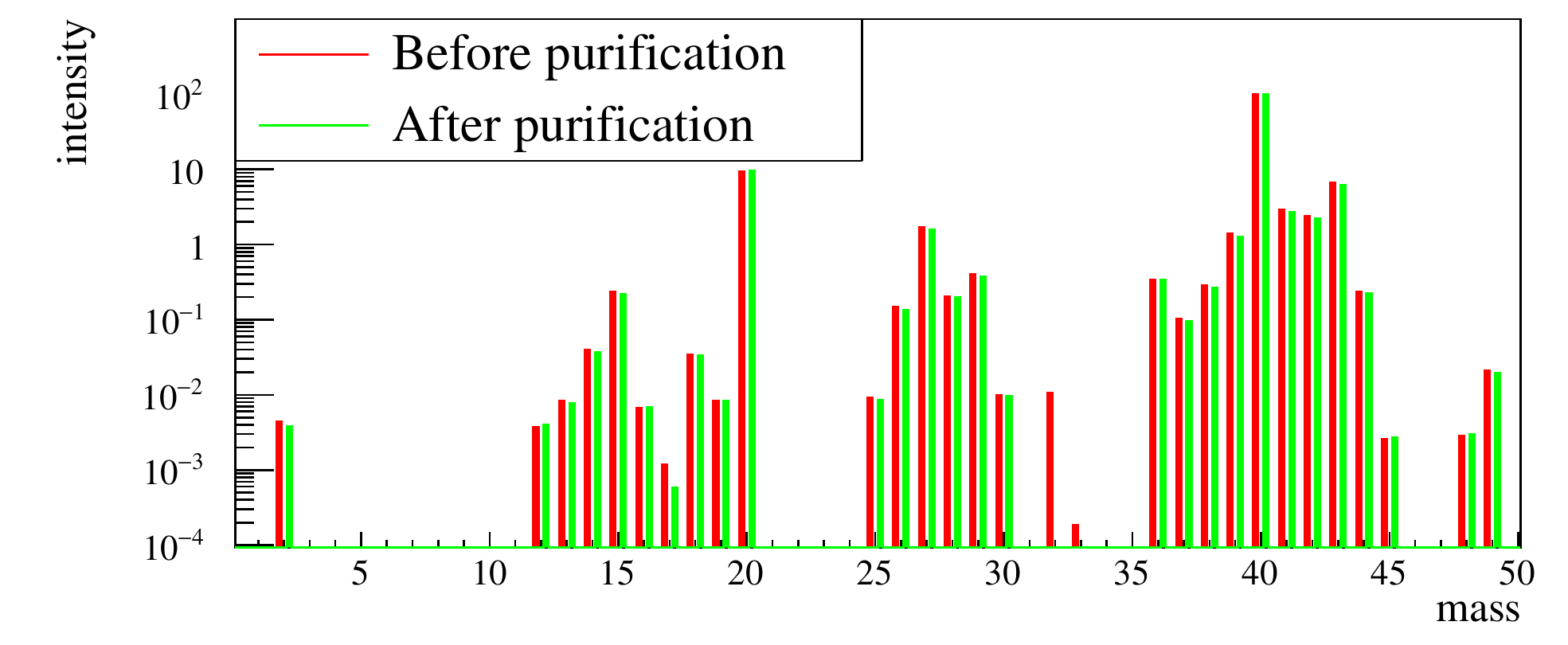}
 \caption{
 Mass spectroscopy results for samples taken before and after purification. 
 The overall composition is similar, except for N=32 (O$_{2}$), that disappeared in the second sample.
 \label{fig:MassSpectrum}
 }
\end{center}
\end{figure}

Gas samples were collected before (\#1, line~3 on
Fig.\ref{fig:fullTime}), and respectively five weeks (\#2) and eight
weeks (\#3) after activating the purification system.
They were analysed in a high resolution ($R=2800$) direct-injection
THERMO MAT 271 mass spectrometer (Fig.~\ref{fig:MassSpectrum},
Table~\ref{tab:MassSpectrum}).
\begin{table} 
\centering
 \footnotesize
\caption{\label{tab:MassSpectrum}
Gas content from mass spectrometry (volume/volume).
Note that the set-up was not sensitive to H$_{2}$O.}
\begin{tabular}{|l|c|c|c|c|c|}
\hline 
\multicolumn{2}{|r|}{Sample \#} & 1 & 2 & 3 \\ 
\multicolumn{2}{|l|}{Compound} & July 08 & Aug. 27 & Sept. 17 \\
\hline
iC$_4$H$_{10}$ & \% & 5.10 & 4.42 & 4.49  \\
O$_2$   & ppm & 180 & $<$20 & $<$20 \\  
CO   & ppm & 190 & 250 & 130 \\  
CO$_2$  & ppm & 120 & 160 & 130 \\  
N$_2$   & ppm & 620 & 890 & 850 \\
\hline
\end{tabular}
\end{table}
The relative amount of O$_2$ and N$_2$ before purification is
consistent with that of air, due to either 
 leaks or outgassing.
Contaminations by CO and CO$_{2}$ probably originate from outgassing
from plastic elements.

After purification, the residual oxygen content is found to be lower
than the detection threshold of the spectrometer.
We also note a small but significant decrease of the amount
of isobutane content upon purification after which it stabilized, 
an effect that is not understood.

\section{Discussion}
\label{sec:discussion}

Garfield++ simulations \cite{Garfield} show that the degradation of
the attenuation during sealed operation, attributed to electron
attachment, is proportional to the oxygen content.
After purification was started the effect disappears exponentially
with a characteristic time $\tau_{\textrm{purif}} = 105\pm4\,\hour$
(Fig.~\ref{fig:fullTime}).
The observed time constant
is larger than the computed value, most likely due to 
the limitation of gas circulation inside the almost closed field cage.
We also observe a decrease of the drift velocity, which cannot be
associated with the oxygen contamination and which is recovered after
purification, and therefore that can be attributed to water contamination.

After purification, the drift velocity was measured to be $6.0 \pm 0.5$\% higher than its initial value.
From the Garfield simulation, this can be explained by the cumulative
effects of the reduced gas pressure from $2.1$ to $1.9\,\bbar$
($+4\%$) and of the measured decrease of the isobutane content from
$5.1$\% to $4.5$\% ($+3\%$).

\section{Conclusion}
\label{sec:conclusion}

We have successfully operated a sealed TPC over five months, with a clear
but manageable gas degradation.
We used the data from cosmic rays in the TPC to monitor the relevant
parameters of the gas (drift velocity, gain, attenuation).
We then purified the gas with a simple, low-power circulation system
and observed a fast and complete recovery of the initial parameters.
If operated routinely, for example on a space mission, the system should
therefore enable a stable performance for several years.

This work is funded by the French National Research Agency
(ANR-13-BS05-0002)

\end{document}